\documentclass[12pt]{article}
\usepackage{epsfig,amssymb,euscript}
\usepackage{amsmath,amscd}

\numberwithin{equation}{section}
\newcommand{\be}{\begin{eqnarray}}
\newcommand{\ee}{\end{eqnarray}}
\newcommand{\bea}{\begin{eqnarray}}
\newcommand{\eea}{\end{eqnarray}}
\newcommand{\ba}{\begin{array}}
\newcommand{\ea}{\end{array}}
\newcommand{\nn}{\nonumber \\}

\newcommand{\bR}{\mathbb{R}}
\newcommand{\bC}{\mathbb{C}}

\def\CG{{\cal G}}

\def\CP{{\cal P}}

\def\cL{{\cal L}}

\def\CQ{{\cal Q}}
\def\CK{{\cal K}}
\def\hto{{\hat{\Omega}}}

\begin{document}

\makeatletter
\renewcommand{\theequation}{\thesection.\arabic{equation}}
\@addtoreset{equation}{section}
\makeatother

\baselineskip 18pt

\begin{titlepage}

\vfill

\begin{flushright}
%Imperial/TP/2008/JG/02\
\end{flushright}

\vfill

\begin{center}
   \baselineskip=16pt
  {\Large\bf HKT Geometry and de Sitter Supergravity}
   \vskip 2cm
      Jai Grover$^{1}$, Jan B. Gutowski$^1$, Carlos A. R. Herdeiro$^2$
      and Wafic Sabra$^3$\\
 \vskip .6cm
      \begin{small}
      $^1$\textit{DAMTP, Centre for Mathematical Sciences\\
        University of Cambridge\\
        Wilberforce Road, Cambridge, CB3 0WA, UK \\
        E-mail: jg372@damtp.cam.ac.uk, J.B.Gutowski@damtp.cam.ac.uk}
        \end{small}\\*[.6cm]
      \begin{small}
      $^2$\textit{Departamento de F\'\i sica e Centro de F\'\i sica do Porto
        , \\ Faculdade de Ci\^encias da Universidade do Porto, \\
        Rua do Campo Alegre, 687, 4169-007 Porto, Portugal \\
        E-mail: crherdei@fc.up.pt}
        \end{small}\\*[.6cm]
      \begin{small}
      $^3$\textit{Centre for Advanced Mathematical Sciences and
        Physics Department, \\
        American University of Beirut, Lebanon \\
        E-mail: ws00@aub.edu.lb}
        \end{small}      
   \end{center}
\vfill

\begin{center}
\textbf{Abstract}
\end{center}

\begin{quote}
Solutions of five dimensional minimal \textit{de Sitter} supergravity admitting
Killing spinors are considered. It is shown that the ``timelike''
solutions are determined in terms of a four dimensional hyper-K\"ahler
torsion (HKT) manifold. If the HKT manifold is conformally
hyper-K\"ahler the most general solution can be
obtained from a sub-class of supersymmetric
solutions of minimal $\mathcal{N}=2$ ungauged supergravity, by means
of a simple transformation. Examples include a
multi-BMPV de Sitter solution, describing multiple rotating black
holes co-moving with the expansion of the universe. If the HKT
manifold is not conformally hyper-K\"ahler, examples admitting a
tri-holomorphic Killing vector field are constructed in
terms of certain solutions of three dimensional Einstein-Weyl geometry.  
\end{quote}

\vfill

\end{titlepage}

%%%%%%%%%%%%%%%%%%%%%%%%%%%%%%%%%%%%%%%%%%%%%%%%%%%%%%%%%%%%%%%%%%%%%%%%%%

%%%%%%%%%%%%%%%%%%%%%%%%%%%%%%%%%%%%%%%%%%%%%%%%%%%%%%%%%%%%%%%%%%%%%%%%%%

\tableofcontents

\section{Introduction}
The connections between complex (in particular K\"ahler) geometry
and supersymmetry have long been known. Well known examples 
arise in the context of
string compactifications; $\mathcal{N}=1$ supersymmetry in
four dimensions requires the compact six dimensional manifold to be a
Calabi-Yau 3-fold \cite{Candelas:1985en}. An even earlier example is
the observation, by Zumino \cite{Zumino:1979et}, that a
two-dimensional non-linear sigma model admits an $\mathcal{N}=2$
supersymmetric extension if and only if the target space metric is
K\"ahler. An $\mathcal{N}=4$ extension requires
a hyper-K\"ahler target space geometry
\cite{AlvarezGaume:1981hm}; essentially each supersymmetry beyond the
first requires the existence of a complex structure. Wess-Zumino-Witten couplings
\cite{Witten:1983ar} in the sigma
model can be interpreted as torsion potentials, from the target space
viewpoint \cite{Curtright:1984dz,Braaten:1985is}. Hence, the inclusion of such couplings leads naturally to
K\"ahler and hyper-K\"ahler torsion (HKT) geometries \cite{howepaphkt, hullhkt}. The latter are
also called heterotic geometries \cite{Delduc:1993pm}, since they
arise in the worldsheet description of soliton solutions of heterotic
string theory \cite{Callan:1991dj}. HKT manifolds
 have also been found in the context of moduli space metrics of 
electrically charged five-dimensional black holes \cite{gibbonshkt, multihkt}.

K\"ahler and hyper-K\"ahler geometry also arise in connection with
supersymmetric solutions in supergravity theories. As a particular
example, all supersymmetric solutions with a timelike Killing vector
field of minimal ungauged \cite{Gauntlett:2002nw}
and gauged five dimensional supergravity \cite{Gauntlett:2003fk} are
defined in terms of a four dimensional base space which is,
respectively, hyper-K\"ahler and K\"ahler. These theories have, as the 
vacuum state, five dimensional Minkowski spacetime and $AdS_5$,
respectively. The purpose of this paper is to show that HKT geometries
also have a role to play in five dimensional supergravities: timelike
solutions (in a sense to be defined below) of \textit{de Sitter}
supergravity \cite{Pilch:1984aw,Lukierski:1984it}
are defined in terms of a base space which is an HKT geometry.

Even though de Sitter superalgebras have only non-trivial representations in a
 positive-definite Hilbert space in two dimensions
\cite{Lukierski:1984it}, the perspective we wish to take here is that
of fake supersymmetry, in analogy to the `Domain Wall/Cosmology' correspondence
\cite{Skenderis:2006jq}: that there is a special class of solutions in
a gravitational theory with a positive cosmological constant admitting
``pseudo Killing'' spinors. Thus, we use supersymmetry as a
solution generating technique. The theory we are considering has,
nevertheless, an interpretation in terms of type IIB* theory \cite{Liu:2003qaa}, which is
related to type IIA string theory via T-duality on a timelike circle \cite{Hull:1998vg}.

This paper is organized as follows. In section 2 we integrate the Killing
spinor equation of minimal five dimensional de Sitter
supergravity. We obtain
the general structure of the solutions of this theory that admit Killing
spinors from which a timelike vector field is constructed. This structure is summarised in section \ref{summary}. In
section 3 we provide some examples and in section 4 we give some final remarks.

\section{Integrating the Killing Spinor Equation}
\subsection{Basic Equations}
\label{basicequations}
The bosonic action is obtained from that in \cite{Gauntlett:2003fk} by
changing the signature and analytically continuing $\chi$. Thus
\[ \mathcal{S}=\frac{1}{4\pi G}\int \left(\frac{1}{4}(^5R-\chi^2)\star
1-\frac{1}{2}F\wedge \star F-\frac{2}{3\sqrt{3}}F\wedge F\wedge A
\right) \ , \]
where $F=dA$ is a $U(1)$ field strength and $\chi\neq 0$ is a real
constant. The
equations of motion are
\be
\label{einsteq}
^5R_{\alpha\beta}-2F_{\alpha\gamma}F_{\beta}^{\
  \gamma}+\frac{1}{3}g_{\alpha\beta}(F^2-\chi^2)=0 \ ,
\ee
and
\be
\label{gaugeq}
 d\star
F+\frac{2}{\sqrt{3}}F\wedge F=0 \ ,  
\ee
where $F^2\equiv F_{\alpha\beta}F^{\alpha\beta}$.

In the minimal theory, the gravitino Killing spinor equation acting
on a Dirac spinor $\epsilon$ is given by

\bea
\label{eqn:grav}
 \bigg[\partial_M +
{1\over4}\Omega_{M,}{}^{N_1 N_2}\Gamma_{N_1 N_2} - {i \over
4\sqrt{3}} F^{N_1 N_2} \Gamma_M \Gamma_{N_1 N_2} +{3i \over
2\sqrt{3}} F_{M}{}^{N} \Gamma_{N} \nonumber \\ + 
\chi({i\over4\sqrt{3}}\Gamma_{M} - {1\over2}A_{M}) \bigg]
\epsilon =0 \ ,
\eea
where $\Omega$ denotes the spin connection.
This Killing spinor equation is obtained from the standard
Killing spinor equation of minimal gauged ($AdS$) D=5 supergravity 
(acting on Dirac spinors) by replacing $\chi \rightarrow i \chi$.
Note that the metric has signature $(-,+,+,+,+)$.

We shall utilize this Killing spinor equation as a solution
generating technique. In particular, if one has a non-vanishing
Killing spinor satisfying ({\ref{eqn:grav}}), and if in addition
the gauge field equations ({\ref{gaugeq}}) are satisfied,
then the integrability conditions of the Killing spinor 
equation place constraints on the Ricci tensor.
For the solutions we consider here, in which the 
Killing spinor generates a timelike vector field,
these constraints on the Ricci tensor are equivalent to
the Einstein equations ({\ref{einsteq}}). This 
can be seen using exactly the same reasoning as in
\cite{Gauntlett:2003fk}. Hence we shall only solve the
Killing spinor and the gauge equations, as the Einstein
equations then follow automatically {\footnote{
However, it should be noted that the same result
does {\it{not}} hold when the Killing spinor generates a null
vector field. In particular, for these solutions,
one component of the Einstein equations must be imposed
in addition to the Killing spinor and gauge equations.
We will not consider such solutions here.}}.

In order to analyse the Killing spinor equations, we
make use of spinorial geometry techniques. 
These were initially used to analyse certain
supersymmetric solutions in ten and eleven-dimensional
supergravity theories.
In the spinorial geometry method, one writes the
spinors as differential forms. Then, by making use
of appropriately chosen gauge transformations
one can transform the spinors to simple canonical
forms, which together with an appropriate choice of basis,
simplifies the analysis.
The result of this is a complete, and systematic, classification of
the different types of spacetime geometry and 
fluxes of supersymmetric solutions.

This method  has been particularly effective in classifying solutions
preserving small and large amounts of supersymmetry
in D=11 and type IIB supergravity \cite{papadopd11}, \cite{papadopiib},
\cite{d11preon}, \cite{iibpreon}. There has also been considerable
progress in the analysis of generic solutions of type I
supergravity using these methods \cite{papadtype1}.
Spinorial geometry techniques have also been 
particularly effective in analysing solutions of lower-dimensional
supergravity theories, for example in \cite{gutpapd4}, \cite{roestd4}, 
\cite{gutnulld5}. Here we apply the same techniques to
analyse the solutions of ({\ref{eqn:grav}}).

For de Sitter supergravity in five-dimensions,
one takes the space of Dirac spinors
to be the space of complexified forms on $\bR^2$, which
are spanned over $\bC$ by $\{ 1, e_1, e_2, e_{12} \}$ where
$e_{12}=e_1 \wedge e_2$. The action of
complexified $\Gamma$-matrices on these spinors is given by

\bea \Gamma_{\alpha} &=& \sqrt{2} e_\alpha \wedge \ ,
\\
\Gamma_{\bar{\alpha}} &=& \sqrt{2} i_{e^\alpha}\ , \eea
for $\alpha=1,2$,
and $\Gamma_0$ satisfies
\be \Gamma_0 1 = -i1 , \quad \Gamma_0 e^{12} = -ie^{12} , \quad
\Gamma_0 e^j = ie^j \ \ j =1,2 \ , \ee
where we work with an oscillator basis in which the spacetime metric is
\be
ds^2 = -({\bf{e}}^0)^2 +2 \delta_{\alpha \bar{\beta}} {\bf{e}}^\alpha {\bf{e}}^{\bar{\beta}} \ .
\ee

\subsection{The timelike case}
In this paper we will focus on integrating the Killing spinor equation
(\ref{eqn:grav}) for the timelike case, i.e. when the vector field
constructed from the Killing spinor is timelike. Note, however, that unlike
the cases of the minimal ungauged \cite{Gauntlett:2002nw} and gauged
theories \cite{Gauntlett:2003fk}, the timelike vector field
obtained from the Killing spinor is \textit{not} a Killing vector
field.  In this case, a generic
spinor can be put in the form $\epsilon = f  1$ by making use
of $Spin(4,1)$ gauge transformations {\footnote{When
the vector field generated from the Killing spinor is null,
then the Killing spinor can be reduced, using gauge transformations,
to the simple canonical form $\epsilon=1+e_1$.}}. Then, we find the
following constraints from the Killing spinor equation

\begin{equation}
{\partial_0f \over f} + {1\over2}\Omega_{0,\mu}{}^\mu -
{1\over2\sqrt{3}}F_\mu{}^\mu + {\chi\over4\sqrt{3}} - {\chi\over2}A_0 =
0 \ ,
\end{equation}

\begin{equation}
\Omega_{0,0\bar{\alpha}} - {2\over\sqrt{3}}F_{0\bar{\alpha}} = 0 \ ,
\end{equation}

\begin{equation}
(\Omega_{0,\bar{\alpha}\bar{\beta}} -
{1\over\sqrt{3}}F_{\bar{\alpha}\bar{\beta}})\epsilon^{\bar{\alpha}\bar{\beta}} = 0 \ ,
\end{equation}

\begin{equation}
{\partial_\alpha f \over f} + {1\over2}\Omega_{\alpha,\mu}{}^\mu +
{3\over2\sqrt{3}}F_{0\alpha} - {\chi\over2}A_\alpha = 0 \ ,
\end{equation}

\begin{equation}
-\Omega_{\alpha,0\bar{\beta}} + \sqrt{3}F_{\alpha\bar{\beta}} -
{1\over\sqrt{3}}F_\mu{}^\mu \delta_{\alpha\bar{\beta}} +
{\chi\over2\sqrt{3}}\delta_{\alpha\bar{\beta}} = 0 \ ,
\end{equation}

\begin{equation}
\Omega_{\alpha,\bar{\mu}\bar{\nu}}\epsilon^{\bar{\mu}\bar{\nu}} +
{2\over\sqrt{3}}F^{0\mu}\epsilon_{\alpha \mu} = 0 \ ,
\end{equation}

\begin{equation}
{\partial_{\bar{\alpha}}f \over f} + {1\over2}\Omega_{\bar{\alpha},\mu}{}^\mu +
{1\over2\sqrt{3}}F_{0\bar{\alpha}} - {\chi\over2}A_{\bar{\alpha}} = 0 \ ,
\end{equation}

\begin{equation}
\Omega_{\bar{\alpha},0\bar{\beta}} - {1\over\sqrt{3}}F_{\bar{\alpha}\bar{\beta}} = 0 \
,
\end{equation}

\begin{equation}
\Omega_{\bar{\alpha},\bar{\mu}\bar{\nu}}\epsilon^{\bar{\mu}\bar{\nu}} = 0 \ .
\end{equation}

However, observe that the Killing spinor equation is invariant under
the $\bR$ transformation for which
\be
\epsilon \rightarrow e^g \epsilon, \qquad A \rightarrow A + {2 \over \chi} dg \ .
\ee

Hence, without loss of generality, we can work in a gauge for which 
$f=1$. This will simplify the analysis of the Killing spinor equation.
In the gauge $f=1$, we obtain the following constraints on the flux components

\bea
\label{fluxv1}
A_0 &=& {1 \over 2 \sqrt{3}} \ ,
\nn 
A_\alpha &=& {1 \over \chi} \Omega_{0,0\alpha} \ ,
\nn
F_{0 \alpha} &=& {\sqrt{3} \over 2} \Omega_{0,0 \alpha} \ ,
\nn
F_{\alpha \beta} &=& \sqrt{3} \Omega_{[\alpha,|0| \beta]} \ ,
\nn
F_{\alpha \bar{\beta}} &=& {1 \over \sqrt{3}} \Omega_{\alpha,0 \bar{\beta}}
+ \big( -{\chi \over 2} + {1 \over \sqrt{3}} \Omega_{\mu,0}{}^\mu \big) 
\delta_{\alpha \bar{\beta}} \ ,
\eea

together with the purely geometric constraints

\be
\label{geo1}
\Omega_{[0, \alpha] \beta} =0, \qquad \Omega_{(\alpha, |0|
\bar{\beta})} = {\chi \over 2 \sqrt{3}} \delta_{\alpha \bar{\beta}},
\qquad
\Omega_{0, \mu}{}^\mu - \Omega_{\mu, 0}{}^\mu + {\chi \over \sqrt{3}}
=0 \ ,
\ee

and

\bea
\label{geo2}
\Omega_{\alpha, \mu \nu} &=& 0 \ ,
\nn
\Omega_{\alpha, \beta}{}^\beta + {1 \over 2} \Omega_{0, 0 \alpha} &=&0 \ ,
\nn
\Omega_{\alpha, \bar{\mu} \bar{\nu}}
-{1 \over 2} \delta_{\alpha \bar{\mu}} \Omega_{0,0 \bar{\nu}}
+ {1 \over 2} \delta_{\alpha \bar{\nu}} \Omega_{0,0 \bar{\mu}} &=&0 \ .
\eea

We begin by analysing the constraints ({\ref{geo1}}). It is convenient to define 
the 1-form $V= {\bf{e}}^0$, and introduce a $t$ co-ordinate such that the dual
vector field is $V= -{\partial \over \partial t}$. Let the remaining (real) co-ordinates be
$x^m$, for $m=1,2,3,4$. The vielbein is then given by

\be
{\bf{e}}^0 = dt + \omega_m dx^m, \qquad {\bf{e}}^{\alpha} = {\bf{e}}^\alpha{}_m dx^m \ .
\ee

It is then straightforward to show that ({\ref{geo1}}) is equivalent to 
\be
(\cL_V {\bf{e}}^\alpha)_{\bar{\beta}} =0 \ ,
\ee
and
\be
(\cL_V {\bf{e}}^\alpha)_\beta = s^\alpha{}_\beta + {\chi \over 2
  \sqrt{3}} \delta^\alpha{}_\beta \ , 
\ee
where 
\be
s^\alpha{}_\beta = \Omega_{0,}{}^\alpha{}_\beta + {1 \over 2} \Omega_{0,\mu}{}^\mu \delta^\alpha{}_\beta
-\Omega_{\beta,}{}^\alpha{}_0 +{1 \over 2} \Omega_{\mu,}{}^\mu{}_0
\delta^\alpha{}_\beta \ ,
\ee
and ({\ref{geo1}}) implies that $s$ is traceless and antihermitian (i.e. $s \in su(2)$).
So, on defining ${\hat{\bf{e}}}^\alpha$ by 
\be
{\bf{e}}^\alpha = e^{-{\chi \over 2 \sqrt{3}} t} {\hat{\bf{e}}}^\alpha
\ ,
\ee
we find
\be
(\cL_V {\hat{\bf{e}}}^\alpha)_\beta = e^{{\chi \over 2 \sqrt{3}}t} s^\alpha{}_\beta \ .
\ee
However, one can without loss of generality apply a $SU(2) \subset Spin(4,1)$ gauge transformation to the Killing
spinors, but which leaves $1$ invariant, and set $s=0$ without loss of generality. 
In this gauge,
\be
\cL_V {\hat{\bf{e}}}^\alpha =0 \ .
\ee

It will be convenient to refer to the 4-manifold with $t$-independent metric 
\be
ds_B^2 = 2 \delta_{\alpha \bar{\beta}} {\hat{\bf{e}}}^\alpha
{\hat{\bf{e}}}^{\bar{\beta}} \ ,
\ee
as the base manifold $B$; the spin connection of this manifold is denoted by ${\hat{\Omega}}$
(whose components will always be taken with respect to the vielbein ${\hat{\bf{e}}}^\alpha$).

Next consider the geometric constraints given in ({\ref{geo2}}).
The third constraint in ({\ref{geo2}}) implies that
\be
\omega  = {2 \sqrt{3} \over \chi} \CP + e^{{\chi \over \sqrt{3}}t} \CQ
\ ,
\ee
where 
\be
\label{pform}
\CP \equiv {\hat{\Omega}}_{{\bar{\beta}}, \alpha}{}^{\bar{\beta}}  {\hat{\bf{e}}}^\alpha
+  {\hat{\Omega}}_{{{\beta}}, {\bar{\alpha}}}{}^{{\beta}}
{\hat{\bf{e}}}^{\bar{\alpha}} = \CP_m dx^m \ ,
\ee
and
\be
\CQ = \CQ_m dx^m \ ,
\ee
are 1-forms on the base manifold $B$ with 
\be
\cL_V \CQ =0 \ .
\ee
Note that, by construction, 
\be
\cL_V \CP =0 \ .
\ee
The remaining geometric content of ({\ref{geo2}}) can be expressed in terms of the spin connection of 
$B$ via
\bea
\label{geo3}
\hto_{\alpha, \mu \nu}=0, \qquad
\hto_{\alpha, \mu}{}^\mu - \hto_{\bar{\mu}, \alpha}{}^{\bar{\mu}} =0 \ ,
\eea
where all components are with respect to the vielbein ${\hat{\bf{e}}}^\alpha$.

Next, consider the constraints on the flux ({\ref{fluxv1}}). The first two constraints fix the gauge potential to be
\be
A = {1 \over 2 \sqrt{3}} (dt + \omega) + {1 \over \sqrt{3}} e^{{\chi \over \sqrt{3}}t} \CQ \ .
\ee
We consider the consistency condition $F=dA$, using the last three constraints in ({\ref{fluxv1}})
to compute $F$, and comparing the resulting expression with $dA$. 
There is no constraint in the ``$0 \alpha$" directions. However, from the $(2,0)$
component of $F=dA$ we obtain
\be
\label{geo4a}
(d \CP)_{\alpha \beta} =0 \ ,
\ee
and from the $(1,1)$ component of $F=dA$, we find
\be
\label{geo4b}
(d \CQ)_{\alpha \bar{\beta}}- {1 \over 2} (d \CQ)_{\mu}{}^\mu \delta_{\alpha \bar{\beta}}
-2 (\CP_\alpha \CQ_{\bar{\beta}} - \CP_{\bar{\beta}} \CQ_\alpha)
+\delta_{\alpha \bar{\beta}} (\CP_\mu \CQ^\mu - \CP_{\bar{\mu}} \CQ^{\bar{\mu}}) =0 \ ,
\ee
together with
\be
\label{geo4c}
(d \CP)_\mu{}^\mu =0 \ ,
\ee
(again, all components are with respect to the vielbein ${\hat{\bf{e}}}^\alpha$).
Observe that ({\ref{geo4b}}) is traceless; in fact, it can be rewritten as
\be
\big( d \CQ -2 \CP \wedge \CQ \big)^+ =0 \ ,
\label{dualprj1}
\ee
where here $+$ denotes the self-dual projection on the base-manifold $B$, with positive orientation 
fixed with respect to the volume form ${\hat{\bf{e}}}^1 \wedge {\hat{\bf{e}}}^{\bar{1}} \wedge {\hat{\bf{e}}}^2 \wedge {\hat{\bf{e}}}^{\bar{2}}$.
Similarly, the constraints ({\ref{geo4a}}) and ({\ref{geo4c}}) on $d \CP$ can be rewritten as
\be
\label{selfdualdp}
(d \CP)^- =0 \ ,
\ee
where $-$ denotes the anti-self-dual projection.
Finally, observe that the constraints ({\ref{pform}}), ({\ref{geo3}}) and ({\ref{selfdualdp}})
are equivalent to
\be
\label{hyperhermit}
d J^i = -2 \CP \wedge J^i \ , \qquad i=1,2,3 \ ,
\ee
where 
\bea
\label{hyperhermit2}
J^1 &=& {\hat{\bf{e}}}^1 \wedge {\hat{\bf{e}}}^2 +  {\hat{\bf{e}}}^{\bar{1}} \wedge  {\hat{\bf{e}}}^{\bar{2}} \ ,
\nn
J^2 &=& i  {\hat{\bf{e}}}^1 \wedge  {\hat{\bf{e}}}^{\bar{1}} +  i  {\hat{\bf{e}}}^2 \wedge  {\hat{\bf{e}}}^{\bar{2}} \ ,
\nn
J^3 &=& -i {\hat{\bf{e}}}^1 \wedge {\hat{\bf{e}}}^2 +i  {\hat{\bf{e}}}^{\bar{1}} \wedge  {\hat{\bf{e}}}^{\bar{2}} \ ,
\eea
defines a triplet of anti-self-dual almost complex structures on $B$ which satisfy the algebra of the imaginary unit
quaternions. 

It should be noted that the constraint ({\ref{hyperhermit}})
 implies that the base $B$ is hyper-K\"ahler with torsion (HKT),
 i.e. 
 \be
 \nabla^+ J^i =0 \ ,
 \ee
 where the connection of the covariant derivative $\nabla^+$ is
 given by
 \be
\Gamma^{(+)}{}^i{}_{jk}=\{{}^i_{jk}\} + H^i{}_{jk}\ ,
\ee
 and where $H$ is the torsion 3-form on $B$ given by
 \be
 H = \star_4 \CP \ .
 \ee
 A HKT manifold is called {\it{strong}} HKT if $H$ is closed. For the solutions under consideration
 here, one can without loss of generality take $B$ to be a strong HKT
 manifold. This is shown in the Appendix.

This exhausts the content of the Killing spinor  equation. Finally we evaluate the gauge
field equations ({\ref{gaugeq}}).
The gauge potential is given by
\be
A = {1 \over 2 \sqrt{3}} dt + {1 \over \chi} \CP + {\sqrt{3} \over 2} e^{{\chi \over \sqrt{3}}t} \CQ \ ,
\label{5dgaugefield}
\ee
with gauge field strength
\be
F = dA = {\chi \over 2} e^{{\chi \over \sqrt{3}}t} {\bf{e}}^0 \wedge \CQ + {1 \over \chi} d \CP + {\sqrt{3} \over 2}
 e^{{\chi \over \sqrt{3}}t}  (d \CQ - 2 \CP \wedge \CQ)  \ .
\label{F}
\ee
The Bianchi identity holds automatically (as the constraints obtained
so far are sufficient to imply $F=dA$).

Note that 
\be
\star F = {\chi \over 2} \star_4 \CQ + {1 \over \chi} {\bf{e}}^0 \wedge d \CP
-{\sqrt{3} \over 2} e^{{\chi \over \sqrt{3}}t} {\bf{e}}^0 \wedge (d \CQ -2 \CP \wedge \CQ) \ ,
\label{starF}
\ee 
 where $\star_4$ denotes the Hodge dual on the 4-dimensional base space $B$
 {\footnote{The five-dimensional volume form $\epsilon^5$ is related to the
 volume form of $B$, $\epsilon^B$ by $\epsilon^5 = e^{-{2 \chi \over \sqrt{3}}t} {\bf{e}}^0 \wedge
 \epsilon^B$, and we use the convention $(\star_4 \CQ)_{i_1 i_2 i_3} =(\epsilon^B){}_{i_1 i_2 i_3}{}^j \CQ_j$,
 where  all indices are with respect to the ${\hat{\bf{e}}}$ vielbein, and are raised/lowered by
 $ds_B^2$.}} .
 
 It is then straightforward to show that the gauge field equations are equivalent to
 \be
 \label{gaugeq1}
 d \star_4 \CQ + {16 \over \sqrt{3} \chi^3} d \CP \wedge d \CP =0 \ .
 \ee
 
\subsection{Summary}
\label{summary}

To summarize, the solutions of the five dimensional theory described
in section \ref{basicequations} are constructed as follows:
\begin{itemize}
\item[1)] Take the base space $B$ to be a four dimensional HKT geometry
  with metric $ds^2_B$ and torsion tensor $H$.
\item[2)] The 1-form $\mathcal{P}$ is given by 
\be
 \mathcal{P}=-\star_4 H \ , 
 \ee
where $\star_4$ denotes the Hodge dual on the base space.
\item[3)] Choose a 1-form $\CQ$ obeying the constraints
  (\ref{dualprj1}) and (\ref{gaugeq1}).
  Note that one can always solve the gauge equation constraint ({\ref{gaugeq1}});
  the general solution is given by
 \be
 \CQ = {16 \over \sqrt{3} \chi^3} \star_4 (\CP \wedge d \CP) + \star_4 d \Phi \ ,
 \ee
 where $\Phi$ is a 2-form on $B$. On substituting this expression back into
 ({\ref{dualprj1}}), one finds an equation constraining $\Phi$, which must be solved.

\item[4)] The spacetime geometry is given by
\be
ds^2 = -\left(dt +  {2 \sqrt{3} \over \chi} \CP + e^{{\chi \over
    \sqrt{3}}t} \CQ\right)^2 + e^{-{\chi \over \sqrt{3}}t} ds_B^2 \ ,
\label{5dmetric}
\ee
where the metric on the base manifold $ds_B^2$ does not depend on $t$,
and $\CP$, $\CQ$ are two $t$-independent 1-forms on $B$. Note that, by
construction,  $B$ also admits three $t$-independent
anti-self-dual almost complex structures $J^i$ which satisfy the algebra of the imaginary unit
quaternions, and also satisfy (\ref{hyperhermit}). It follows that
(\ref{selfdualdp}) is obeyed.
\item[5)] The gauge
potential is given by (\ref{5dgaugefield}).
\end{itemize}

Note that the Ricci scalar of the five dimensional metric is given by 
\be
\mathcal{R}=\frac{F_{\mu\nu}F^{\mu\nu}+5\chi^2}{3} \ ; 
\ee
using (\ref{F}) and (\ref{starF}) we find that
\be
\begin{array}{l}
\displaystyle{\mathcal{R}=\frac{5}{3}\chi^2+\frac{e^{\frac{2\chi}{\sqrt{3}} t}}{3}\left[\frac{(d\CP)^2}{\chi^2}-\frac{\chi^2}{2}e^{\frac{\chi}{\sqrt{3}} t}\CQ^2  +\frac{3}{4}e^{\frac{2\chi}{\sqrt{3}} t}(d\CQ-2\CP\wedge \CQ)^2\right]} \ , \end{array}
\ee
where the norms are computed with respect to the $t$-independent 
base space metric. Therefore, the $t$-dependence of
the Ricci scalar can be read directly from the above expression.
Thus, in particular, for the solution to be regular at both
$t=\pm \infty$ we must require
\be
 \CQ=0 \ , \qquad d\CP=0 \ . 
 \ee
In particular, this implies that the base space is conformally hyper-K\"ahler.

 \section{Examples}
  
 In this section, we present some examples of solutions.

 \subsection{Solutions with conformally hyper-K\"ahler base}
 
 Suppose that the base space $B$ is conformal to
 a hyper-K\"ahler manifold $HK$.
 We set ${\bf{e}}^\alpha = e^\phi {\bf{E}}^\alpha$,
 where $\phi$ is a real $t$-independent function, such that
 the manifold $HK$ with metric
 \be
 ds_{HK}^2  = 2 \delta_{\alpha \bar{\beta}} {\bf{E}}^\alpha {\bf{E}}^{\bar{\beta}} \ ,
 \ee
 is hyper-K\"ahler with closed K\"ahler forms ${\tilde{J}}^i$
 related to $J^i$ by
 \be
 J^i = e^{2 \phi} {\tilde{J}}^i \ .
 \ee
 
 Then it is straightforward to show that ({\ref{hyperhermit}})
 implies that
 \be
 \CP = - d \phi \ ,
 \ee
 so in particular, $d \CP=0$. Also, ({\ref{dualprj1}})
 is equivalent to 
 \be
 d( e^{2 \phi} \CQ)^+ =0 \ ,
\ee
where here $+$ denotes the self-dual projection on $HK$,
and the gauge equation constraint ({\ref{gaugeq1}}) is equivalent
to
\be
d (e^{2 \phi} \star_{HK} \CQ)=0 \ ,
\ee
 where $\star_{HK}$ denotes the Hodge dual on $HK$. 
 
 The solution can be simplified further by making the co-ordinate transformation
 \be
 t = t' + {2 \sqrt{3} \over \chi} \phi \ ,
\ee
and setting
\be
\CG = e^{2 \phi} \CQ \ .
\ee

The metric is then given by
\be
ds^2 = -(dt' + {e^{{\chi \over \sqrt{3}}t'}} \CG)^2 +  {e^{-{\chi \over \sqrt{3}}t'}} ds_{HK}^2 \ ,
\ee
where
\be
\label{simplerhk}
(d \CG)^+ =0 , \qquad d \star_{HK} \CG =0 \ .
\ee
In these co-ordinates, the gauge potential and gauge field strength are
\be
A = {1 \over 2 \sqrt{3}} dt' + {\sqrt{3} \over 2}  {e^{{\chi \over \sqrt{3}}t'}} \CG \ ,
\qquad 
F = {\sqrt{3} \over 2} d ( {e^{{\chi \over \sqrt{3}}t'}} \CG) \ .
\ee

We can put this class of solutions in a more familiar form. Set
\be 
\CG=\frac{\sqrt{3}}{\chi}dV'+a'  \ . 
\ee
Making a coordinate transformation $t'\rightarrow t''$ given by
\be
  t''=\frac{\sqrt{3}}{\chi}\left(V'-e^{-\frac{\chi}{\sqrt{3}}t'}\right) \ , 
\ee  
the solution can the be written in the form
\be ds^2=-f^2(dt''+a')^2+f^{-1}ds^2_{HK} \ , \ \ \ \
F=\frac{\sqrt{3}}{2}d \bigg( f(dt''+a') \bigg) \ , \ee
where 
\be f^{-1}=V'-\frac{\chi}{\sqrt{3}}t'' \ ,   \ee
and 
\be \Delta_{HK} V'=-\frac{\chi}{\sqrt{3}}\nabla_{HK} \cdot a' \ , \qquad
(d a')^+ =0 \ , 
\ee
where $\nabla_{HK} \cdot a'$ is the covariant divergence of $a'$ in
$HK$. Finally we can choose a Lorentz-type gauge for $a'$: setting 
\be
a'=a+d\zeta \ , \qquad  t''+\zeta=t \ , \qquad \
V=V'+\frac{\chi}{\sqrt{3}}\zeta \ , 
\ee
and choosing $\zeta$ such that $\nabla_{HK} \cdot a=0$ we find the final form
\be ds^2=-f^2(dt+a)^2+f^{-1}ds^2_{HK} \ , \ \ \ \
F=\frac{\sqrt{3}}{2}d \bigg( f(dt+a)  \bigg) \ , \label{usualform}\ee
where 
\be f^{-1}=V-\frac{\chi}{\sqrt{3}}t \ ,  \ \label{f}\ee
and 
\be \Delta_{HK} V=0 \ , \qquad
(d a)^+ =0 \ . \label{conditionscanonical}
\ee
$dS_5$ is obtained by taking $HK=\mathbb{R}^4$, $V=const.$ and
$a=0$.

Observe that in the limit of zero cosmological constant, the solutions
(\ref{usualform})-(\ref{conditionscanonical}) become a subset of the supersymmetric solutions of the
ungauged theory \cite{Gauntlett:2002nw}, namely those with a timelike
Killing vector field and $G^+=0$ (in
the notation therein). Any such solution - which we call \textit{seed}
solution - can be made into an asymptotically de
Sitter solution, simply by \textit{adding} a linear time dependence to
the harmonic function of the seed solution. Such a procedure for building $dS$ solutions was
first observed in \cite{Behrndt:2003cx} and it underlies several
asymptotically dS solutions constructed in the last few years. However, note that
the result presented here is stronger than the result presented in  \cite{Behrndt:2003cx}:
\bigskip

\textit{Any solution of (\ref{einsteq}), (\ref{gaugeq}) with a supercovariantly constant
  spinor and a base space which is conformal to a hyper-K\"ahler
  manifold is of the form
  (\ref{usualform})-(\ref{conditionscanonical}). Thus it can be
  obtained from a seed solution of the $\mathcal{N}=2$, $D=5$ minimal
  ungauged supergravity theory with $G^+=0$ simply by adding a linear time
  dependence to the harmonic function, as in (\ref{f}).}

\bigskip

 For instance, the multi-centred, non-rotating, black hole solutions of
\cite{liusabra} are obtained taking $HK=\bR^4$ and $a=0$; i.e. the
five dimensional Majumdar-Papapetrou multi-black hole solution is the
seed. Introducing rotation, one finds solutions whose seeds are the
BMPV black hole \cite{Breckenridge:1996is}, G\"odel type universes
\cite{Gauntlett:2002nw} or black holes in G\"odel type universes
\cite{Gauntlett:2002nw,Herdeiro:2002ft}. To be concrete let us analyse a
single-centred solution with rotation. 

Set $HK=\mathbb{R}^4$ written in terms of left (or right)
invariant forms on $SU(2)$:
\bea
ds^2({\mathbb{R}}^4)&=&dr^2+\frac{r^2}{4}\left((\sigma_L^1)^2+(\sigma_L^2)^2+(\sigma_L^3)^2\right)
\nn
&=&dr^2+\frac{r^2}{4}\left((\sigma_R^1)^2+(\sigma_R^2)^2+(\sigma_R^3)^2\right)
\ . 
\eea
An explicit expression for the 1-forms $\sigma_{L,R}$ in terms of Euler
angles can be found, for instance, in \cite{Gauntlett:2002nw}. Let $V$
be harmonic on $\mathbb{R}^4$. Set 
\be 
a=g_i^L(r)\sigma_L^i+g_i^R(r)\sigma_R^i \ . 
\ee
Noting that 
\be d\sigma_L^i=-\frac{1}{2}\epsilon^{ijk}\sigma_L^j\wedge\sigma_L^k \ ,
\ \ \ \
d\sigma_R^i=\frac{1}{2}\epsilon^{ijk}\sigma_R^j\wedge\sigma_R^k \ , 
\ee
it follows that
the equations
(\ref{conditionscanonical}) are obeyed if
\be 
g_i^L(r)=C^L_i r^2 \ , \ \ \ \ \  \ \ g_i^R(r)=\frac{C^R_i}{r^2} \ , 
\ee
where $C_i^{L,R}$ are constants. For a particular choice of the constants, $C^R_i=0=C_3^L$, this is a
solution dubbed `G\"odel-de-Sitter Universe' in
\cite{Behrndt:2004pn}. If $V=1$, the seed is the maximal
supersymmetric G\"odel Universe found in \cite{Gauntlett:2002nw}; but
if $V=1+\mu/r^2$ the seed is actually the G\"odel universe black hole
found in  \cite{Gauntlett:2002nw} and discussed in
\cite{Herdeiro:2002ft}. On the other hand, taking $V=1+\mu/r^2$, $C_i^L=0=C_1^R=C_2^R$,
$C_3^R=j$; then
\be f^{-1}=1+\frac{\mu}{r^2}-\frac{\chi}{\sqrt{3}}t \ , \ \ \
a=\frac{j}{r^2}\sigma_{R}^3 \ . \label{bmpv} \ee
The seed is now the 
BMPV black hole \cite{Breckenridge:1996is}. Thus we dub this solution
a BMPV-de Sitter black hole. It was first found, albeit not in this
form, in \cite{Klemm:2000gh}. This solution can be easily generalised to a
multi-BMPV-de-Sitter solution by taking as seed the multi-centred BMPV solution
 \cite{Gauntlett:1998fz}; using Cartesian coordinates on $\mathbb{R}^4$
 we have 
\be f^{-1}=1+\sum_i \frac{\mu_i}{|{\bf x} - {\bf x}^i|^2} -\frac{\chi}{\sqrt{3}}t \ , \ \ \
a=dx^iJ_i^{\ k}\partial_k \left(\sum_l \frac{j_l}{|{\bf x} - {\bf x}^l|^2}\right) \ , \label{multibmpv} \ee
where $J$ is a complex structure on $\mathbb{R}^4$ and $\mu_i,j_i$ and
${\bf x}^i$ are
constants. 

Let us note that the multi-black
hole solutions displayed in this section are a five dimensional
generalisation of the Kastor-Traschen solutions
\cite{Kastor:1992nn}, but which can also carry rotation. 
Analogous solutions for branes
have been studied in \cite{Chen:2005jp,Gibbons:2005rt}.

Finally let us remark that the solutions of the ungauged supergravity
theory with $G^+\neq 0$ do not generalise straightforwardly to the de
Sitter case. Most notably this includes the supersymmetric black ring
of \cite{Elvang:2004rt}.

\subsection{Solutions with a tri-holomorphic Killing vector}

Suppose that the strong HKT base manifold $B$ has a
tri-holomorphic Killing vector $X$, such that
\be
\cL_X h_B=0, \qquad \cL_X J^i =0, \qquad i=1,2,3 \ ,
\ee
where $h_B$ is the metric on $B$.
Such manifolds have been classified in \cite{chavetodvalent}, \cite{gaudtod},
and their structure is specified in terms of a constrained 3-dimensional Einstein-Weyl
geometry. 
This consists of a 3-dimensional manifold $E$ equipped with a metric $\gamma_{ij}$,
together with a 1-form $u$ on $E$, and a scalar $u_0$ which satisfy the constraints
\be
\label{einsteinweyl1}
\star_E du = -d u_0 - u_0 u \ ,
\ee
and
\be
\label{einsteinweyl2}
{}^{(E)}R_{ij} + \nabla_{(i} u_{j)} + u_i u_j = \gamma_{ij} ({1 \over 2} u_0^2 + u_k u^k) \ ,
\ee
where ${}^{(E)}R_{ij}$ denotes the Ricci curvature of the Levi-Civita connection of $E$,
denoted here by $\nabla_i$.
Furthermore, $\star_E$ denotes the Hodge dual of $E$. The $1$-form $u$ is 
also co-closed
\be
\label{einsteinweyl3}
d \star_E u =0 \ .
\ee

Then the 4-dimensional base geometry is obtained by introducing
a local co-ordinate $\tau$ such that $X={\partial \over \partial \tau}$.
We assume that $X$ is a symmetry of the full five-dimensional solution.
The metric on $B$ is
\be
\label{tribasemet}
ds_B^2 = {1 \over W} (d \tau + \Psi)^2 + W ds^2_E \ ,
\ee
where $W$ is a $\tau$-independent function, and $\Psi$ is
$\tau$-independent 1-form on the Einstein-Weyl manifold $E$
which are related by the constraint
\be
\label{tri1a}
\star_E d \Psi = d W + W u \ .
\label{solvepsi}
\ee
The scalar $u_0$, 1-form $u$ and metric on $\gamma_{ij}$ on $E$ do not depend on $\tau$.

Observe that the constraints ({\ref{einsteinweyl1}}), ({\ref{einsteinweyl3}}), ({\ref{tri1a}})
imply that
\be
(\Delta_E + u^i \nabla_i)W = 0, \qquad (\Delta_E + u^i \nabla_i) u_0
=0 \ ,
\label{Wconstraint}
\ee
where $\Delta_E$ is the Laplacian on $E$. The volume form on $B$ and the
volume form on $E$, ${\rm \ dvol}_{E}$ are related via
\be
\epsilon^B = W (d \tau + \Psi) \wedge {\rm \ dvol}_{E} \ .
\ee 

The torsion is obtained using the identification
\be
\CP = -{u_0 \over 2 W} (d \tau + \Psi)-{1 \over 2}u \ ,
\label{curlyp}
\ee
and it is straightforward to verify that $\CP$ is co-closed on $B$,
so the geometry is indeed strong HKT.

To proceed, consider the constraint ({\ref{dualprj1}}), and write
\be
\CQ = \CQ_\tau (d \tau + \Psi) + {\tilde{\CQ}} \ ,
\label{curlyq}
\ee
where $\CQ_\tau$ is a $\tau$-independent function, and
$\tilde{\CQ}$ is a $\tau$-independent 1-form on $E$.
It is then straightforward to show that 
({\ref{dualprj1}}) is equivalent to
\be
\label{triaux1}
d {\tilde{\CQ}} + u \wedge {\tilde{\CQ}} + \star_E (Q_\tau dW - W d Q_\tau) + u_0 \star_E {\tilde{\CQ}} =0 \ .
\ee

Next consider
({\ref{gaugeq1}}), this can be rewritten as
\be
d \star_E {\tilde{\CQ}} = {8W \over \sqrt{3} \chi^3} d({u_0 \over W}) \wedge \star_E d({u_0 \over W})  \ .
\ee

By making use of ({\ref{triaux1}}), together with the other constraints this equation can be rewritten as
\be
(\Delta_E + u^i \nabla_i) \bigg( \CQ_\tau - {4 \over 3 \sqrt{3} \chi^3 }{u_0^3 \over W^2} \bigg) =0 \ ,
\ee
and hence
\be 
\label{cqsol1}
\CQ_\tau = {4 \over 3 \sqrt{3} \chi^3 }{u_0^3 \over W^2} +M \ ,
\label{curlyqt}
\ee
where $M$ satisfies 
\be
(\Delta_E + u^i \nabla_i) M =0 \ .
\label{Mconstraint}
\ee

On substituting this expression back into
({\ref{triaux1}}) and defining $\CK$ by
\be
{\tilde{\CQ}} = {4 \over \sqrt{3} \chi^3} d({u_0^2 \over W}) + {4 \over \sqrt{3} \chi^3} {u_0^2 \over W} u  + \CK \ ,
\label{curlytq}
\ee
we then obtain the constraint
\be
\label{triaux2}
d \CK + u \wedge \CK + \star_E (M dW - W dM) + u_0 \star_E \CK =0 \ .
\ee

Observe that this constraint implies  that
\be
d \star_E \CK =0 \ .
\ee

To summarise, solutions with a tri-holomorphic Killing vector field
are constructed in the following way:
\begin{itemize}
\item[1)] Choose the 3-dimensional Einstein-Weyl data,
  $(\gamma_{ij},u_i,u_0)$, which must satisfy (\ref{einsteinweyl1}),
  (\ref{einsteinweyl2}) and (\ref{einsteinweyl3}).
\item[2)] Choose a function $W$ satisfying (\ref{Wconstraint}).
\item[3)] Solve (\ref{solvepsi}) to obtain the 1-form $\Psi$; $\CP$ is
  then obtained from (\ref{curlyp}) and the base space from (\ref{tribasemet}).
\item[4)] Choose a function $M$ satisfying (\ref{Mconstraint}).
  $\CQ_{\tau}$ is then obtained from (\ref{curlyqt}).
\item[5)] Choose a 1-form $\CK$ obeying (\ref{triaux2}); $\tilde{\CQ}$
  is then obtained from (\ref{curlytq}) and $\CQ$ from (\ref{curlyq}).
\item[6)] The five dimensional metric and gauge field are obtained
  from (\ref{5dmetric}) and (\ref{5dgaugefield}).
\end{itemize}

\subsubsection{Example: The Round Sphere}
A basic example for which the base space is \textit{not} conformally
hyper-K\"ahler ($d\CP\neq 0$) is obtained by taking the Einstein-Weyl data to be
\be
 ds_E=b^2\left(d\theta^2+\sin^2\theta(d\phi^2+\sin^2\phi d\psi^2)\right)
\ , \ \ \ \ u=0 \ , \ \ \ \ u_0=-\frac{2}{b} \ , 
\ee
which has been considered in \cite{Delduc:1993pm} and \cite{papadelipmon}.
Following the algorithm explained above we choose
\be 
W=\alpha\cot \theta \ , 
\ee
where $\alpha$ is a constant, and obtain
\be
 \Psi=b\alpha\cos\phi d\psi \ , \ \ \ \ \ \
\CP=\frac{1}{b\alpha\cot\theta}\left(d\tau+b\alpha\cos\phi d\psi\right)
\ ;
\ee
the base space is then
\be
ds^2_B=\frac{1}{\alpha\cot\theta}\left(d\tau+b\alpha\cos\phi
  d\psi\right)^2+b^2\alpha\cot\theta
  \left(d\theta^2+\sin^2\theta(d\phi^2+\sin^2\phi d\psi^2)\right) \ . 
 \ee 
Now we choose
\be 
M=\beta_1+\beta_2\cot\theta \ ,
\ee
then
\be
\CQ_{\tau}=-\frac{32\tan^2\theta}{3\sqrt{3}b^3\alpha^2\chi^3}+\beta_1+\beta_2\cot\theta
\ . 
\ee

To solve (\ref{triaux2}) we take $\CK=dV$, where $V$ is a function; then (\ref{triaux2}) is
solved by taking
\be 
\CK=\frac{b\alpha\beta_1}{2} \, d\cot\theta \ , 
\ee
and hence
\be
\tilde{\CQ}=d\left(\frac{16\tan\theta}{\sqrt{3}b^2\alpha\chi^3}+\frac{b\alpha\beta_1
\cot\theta}{2} \right) \ . 
\ee
Rescaling the coordinate $\tau \rightarrow b\alpha \tau$ and
introducing a radial coordinate $R$ by\footnote{Note that $R$ has
  dimensions $[R]=L^2$.}
\be
 R\chi^2=\tan\theta \ , 
\ee 
 the five dimensional metric can be written
\be ds^2_5=-\left(dt'+2\sqrt{3}\chi R\left(d\tau+\cos\phi
  d\psi\right)+e^{\frac{\chi}{\sqrt{3}}t'}\CQ\right)^2+e^{-\frac{\chi}{\sqrt{3}}t'}ds^2_B
  \ ,
  \ee
where the base space is
\be ds^2_B=R\left(d\tau+\cos\phi
  d\psi\right)^2+\frac{1}{R(1+R^2\chi^4)}
  \left(\frac{dR^2}{1+R^2\chi^4}+R^2(d\phi^2+\sin^2\phi d\psi^2)\right)  \label{newbase}\ee 
and $\CQ$ is given by
\be
\CQ=\left(\frac{\sqrt{3}\mu\chi}{2}-\frac{32\chi^3}{3\sqrt{3}}R^2+\frac{j}{4R}\right)(d\tau+\cos\phi
d\psi)+d\left(\frac{16\chi}{\sqrt{3}}R+\frac{\sqrt{3}\mu}{4\chi R}\right) \ . 
\ee
We have introduced $\mu\equiv 2b^3\alpha^2\beta_1\chi/\sqrt{3}$,
$j\equiv 4b^3\alpha^2\beta_2$. Note also that
we have shifted the time coordinate, $t=
t'+\frac{\sqrt{3}}{\chi}\ln (b^2\alpha\chi^2)$. Observe that for
$\chi=0$, the base space (\ref{newbase})  is Euclidean 4-space written in a
Gibbons-Hawking form \cite{Gibbons:1979zt}.

Finally, introducing a new radial coordinate $r$ by
\be
R=\frac{r^2}{4} \ , 
\ee
and a new time coordinate $t$ by
\be
t=\frac{4\chi}{\sqrt{3}}r^2+\frac{\sqrt{3}\mu}{\chi
  r^2}-\frac{\sqrt{3}}{{\chi}}e^{-\frac{\chi}{\sqrt{3}}t'} \ , \ee
the metric is written
\be
ds^2_5=-f^2\left[dt+\left(\frac{4\chi^3}{3\sqrt{3}}r^4+\sqrt{3}\mu\chi+\frac{j}{r^2}-\frac{\chi^2}{2}tr^2\right)\sigma_3\right]^2+f^{-1}ds^2_B
  \ , \label{bmpvtorsionmetric}\ee
where
\be f^{-1}\equiv
  \frac{4\chi^2}{3}r^2+\frac{\mu}{r^2}-\frac{\chi}{\sqrt{3}}t \ , \ee
and the base space is
\be ds^2_B=\frac{1}{1+\left(\frac{r\chi}{2}\right)^4}
  \left(\frac{dr^2}{1+\left(\frac{r\chi}{2}\right)^4}+\frac{r^2}{4}(\sigma_1^2+\sigma_2^2)\right)+\frac{r^2}{4}\sigma_3^2 \ , \label{newbase2}\ee 
where we have used the following (right) one forms on $SU(2)$
\bea
\sigma_1 &=& \cos \tau \sin \phi d \psi - \sin \tau d \phi \ ,
\nn
\sigma_2 &=& \sin \tau \sin \phi d \psi + \cos \tau d \phi \ ,
\nn
\sigma_3 &=& \cos \phi d \psi + d \tau \ .
\eea
The gauge field strength for this solution is
\be
F=\frac{\sqrt{3}}{2}d \bigg( f\left[dt+\left(\frac{2\mu\chi}{\sqrt{3}}+\frac{j}{r^2}-\frac{\chi^2}{6}tr^2\right)\sigma_3\right]\bigg)
\ . \label{bmpvtorsiongauge}\ee

To consider two limits of the solution
(\ref{bmpvtorsionmetric})-(\ref{bmpvtorsiongauge}) it is convenient to
shift
\be
 t \rightarrow t-\frac{\sqrt{3}}{\chi}c \ . 
 \ee
Then, observe that for $\chi=0$ the solution is the BMPV
black hole.  For small $r$, the
dominating terms are also the ones of the BMPV black hole. Thus, this
geometry contains a BMPV black hole. Note also that performing the
rescalings
\be 
(t,r,c) \rightarrow
\left(\frac{t}{\eta^2},r\eta,\frac{c}{\eta^2}\right) \ , 
\ee
and taking the
limit $\eta\rightarrow 0$ one recovers the solution (\ref{bmpv}).

Let us note that the solution just derived is singular. Using the
$t',R$ coordinates one verifies that, in accordance with the comments
at the end of section (\ref{summary}), the solution has a curvature
singularity at $t'\rightarrow +\infty$ (and also at $R\rightarrow
+\infty$). It remains to be seen if the singularities that will
necessarily arise in the non conformally hyper-K\"ahler case can have
interesting interpretations as Big Bang/Big Crunch singularities or
black object singularities.

\section{Final Remarks}
In this paper we have shown that the timelike solutions of five
dimensional, minimal de
Sitter supergravity admitting Killing spinors are determined by a four
dimensional HKT geometry, wherein two constraint equations have to be
solved, as summarised in section \ref{summary}. To give concrete
examples we considered two distinct cases:

\begin{itemize}

\item When the HKT manifold is conformally hyper-K\"ahler,
  all solutions can be  generated from supersymmetric solutions (with a timelike
  Killing vector field) of five dimensional, minimal, ungauged supergravity,  in the following way: Take a solution with
  $G^+=0$ (in the notation of \cite{Gauntlett:2002nw}) and add a
  linear time dependence with the appropriate coefficient (\ref{f}) to the
  harmonic function of the solution. Our analysis shows that
  all solutions with a conformally hyper-K\"ahler base space can be
  put in this form, which is therefore a stronger statement than that
  of  \cite{Behrndt:2003cx}. Several examples were given, including a
  multi-BMPV de Sitter solution, describing multiple rotating black
  holes co-moving with the expansion of the universe.

\item If one assumes  that the HKT manifold is {\it{not}} conformally
  hyper-K\"ahler, but possesses a tri-holomorphic
  Killing vector field, solutions can be found in terms of certain
  constrained (special, in the terminology of \cite{chavetodvalent})
  three-dimensional Einstein-Weyl geometries. Taking the latter to be
  simply the round three-sphere an explicit example was constructed,
  describing a BMPV black hole inside a singular universe.

\end{itemize}

One immediate task that this work suggests is to look for more interesting
solutions in the non-conformally hyper-K\"ahler case. As discussed in
section \ref{summary}, such solutions will always have curvature
singularities at $t=\pm \infty$, but these might have a cosmological
or black object interpretation. 

It would be particularly interesting to determine whether there
exist regular (pseudo) supersymmetric black ring solutions in de Sitter supergravity.
One encouraging hint that such ring geometries may exist is the fact that certain
solutions, such as multi-BMPV black holes, have been found in both the ungauged and
the de Sitter supergravities. In contrast to this, there are no known
supersymmetric, asymptotically $AdS_5$ multi-black hole solutions.
However, it is not possible to straightforwardly construct a de-Sitter
black ring using the asymptotically flat solution found in
\cite{Elvang:2004rt} as a {\it seed},
as one can do for the (multi) BMPV black hole.
This is because the asymptotically flat ring solution has $G^+ \neq 0$.
Therefore if a (pseudo) supersymmetric black ring exists
in this theory it will be described by a base space which is not
conformally hyper-K\"ahler.

One other issue which remains to be resolved is whether
one can always construct a 5-dimensional solution
given a {\it generic} HKT base space $B$. In 
particular, it is not a priori apparent that given $B$, one can 
always find a solution to both the constraints
(\ref{dualprj1}) and (\ref{gaugeq1}), although we have shown that
one can always solve ({\ref{gaugeq1}}). Note that
in the case of the $AdS$ supergravity theory, it
was shown in \cite{Figueras:2006xx} that not all K\"ahler bases give
rise to a five dimensional solution.

Finally, an immediate continuation of this work consists of
considering the null case. Other possible generalisations include going
beyond the minimal theory by including vector multiplets, and by
considering de Sitter supergravity in other dimensions.

\section*{Acknowledgments}
We are very grateful to G. Gibbons, G. Papadopoulos and A. Swann for discussions
concerning HKT geometry. C.H. and W.S. would like to thank the hospitality of
DAMTP Cambridge, where part of this work
was done. Centro de F\'\i sica do Porto is partially funded by FCT
through the POCI programme. The work of W.S. is supported in
part by the National Science Foundation under grant number PHY-0703017.

\renewcommand{\theequation}{A-\arabic{equation}}
  % redefine the command that creates the equation no.
  \setcounter{equation}{0}  % reset counter 
  \section*{Appendix: Strong HKT manifolds}  % use *-form to suppress numbering

For the solutions under consideration
 here, one can without loss of generality take $B$ to be a strong HKT manifold.
 This can be achieved by making a conformal transformation, and defining
 a new vielbein ${\bf{E}}^\alpha$ by
 \be
 {\bf{{\hat{e}}}}^\alpha = e^\phi {\bf{E}}^\alpha \ ,
 \ee
 where $\phi$ is a $t$-independent real function. 
 Then the manifold $B'$ with
 metric
  \be
 ds_{B'}^2  = 2 \delta_{\alpha \bar{\beta}} {\bf{E}}^\alpha {\bf{E}}^{\bar{\beta}} \ ,
 \ee
 admits forms ${\tilde{J}}^i$
 related to $J^i$ by
 \be
 J^i = e^{2 \phi} {\tilde{J}}^i \ ,
 \ee
which satisfy the algebra of the imaginary unit quaternions, and
\be
 d {\tilde{J}}^i = -2 \CP' \wedge {\tilde{J}}^i \ ,
 \ee
 where 
 \be
 \CP' = \CP+d \phi \ .
 \ee
Hence $B'$ is also a HKT manifold, and by making an appropriate
choice of $\phi$, one can ensure that 
\be
d \star'_4 \CP' =0 \ ,
\ee
where $\star'_4$ denotes the Hodge dual on $B'$.
This implies that the torsion $H'= \star'_4 \CP' $ is closed.
With this choice, $B'$ is strong HKT.
Furthermore, on defining 
\be
\CQ' = e^{2 \phi} \CQ, \qquad t'=t-{2 \sqrt{3} \over \chi} \phi \ ,
\ee
one finds that the metric, gauge potential and the 
constraints ({\ref{dualprj1}}) and 
({\ref{gaugeq1}})  remain invariant, with $t$, $\star_4$, $\CP$ and $\CQ$
replaced with $t'$, $\star'_4$, $\CP'$ and $\CQ'$.

Hence, one can without loss of generality drop the primes, and take the base manifold $B$
to be strong HKT.

\end{document}